\documentclass[aps,prr,longbibliography,twocolumn]{revtex4-2}
\usepackage[T1]{fontenc}
\usepackage[latin9]{inputenc}
\usepackage{amsmath}
\usepackage{graphicx}
\usepackage{babel}
\usepackage{hyperref}
\begin{document}
\title{Eigenstate thermalization to non-monotonic distributions in strongly-interacting chaotic lattice gases}
\author{Vladimir A. Yurovsky}
\email{volodia@post.tau.ac.il}{
\affiliation{School of Chemistry, Tel Aviv University, 6997801 Tel Aviv, Israel}
\author{Amichay Vardi}
\email{avardi@bgu.ac.il}
\affiliation{Department of Chemistry, Ben-Gurion University of the Negev, Beer-Sheva
84105, Israel}
\affiliation{ITAMP, Harvard-Smithsonian Center for Astrophysics, Cambridge, MA 02138, USA}
\date{\today}

\begin{abstract}
We find non-monotonic equilibrium energy distributions, qualitatively different from the Fermi-Dirac and Bose-Einstein forms, in strongly-interacting many-body chaotic systems. The effect emerges in systems with finite energy spectra, supporting both positive and negative temperatures, in the regime of quantum ergodicity. The results are supported by  exact diagonalization calculations for chaotic Fermi-Hubbard and Bose-Hubbard models, when they have Wigner-Dyson statistics of energy spectra and demonstrate eigenstate thermalization. The proposed effects may be observed in experiments with cold atoms in optical lattices.
\end{abstract}
\maketitle
\section{Introduction}
\label{sec:intro}
The properties of complex systems in thermodynamic equilibrium are determined
by a few  thermodynamic parameters, such as temperature, pressure,
and density. Chaotic systems relax to equilibrium independently
of their specific initial state. However, an isolated quantum system
is described by the Schr\"{o}dinger equation.
Launching such a system in one of its eigenstates, it
would remain in that state forever  and the expectation value of any observable would be constant. This seems to violate the existence of a thermodynamic equilibrium state into which the system relaxes independently of its initial preparation.

The paradox is
resolved by the eigenstate thermalization hypothesis (ETH) \cite{deutsch1991,srednicki1994}
(see also \cite{rigol2008,khodja2015}, the experimental work \cite{kaufman2016},
the review \cite{deutsch2018} and the references therein). It states
that the vast majority of a chaotic system's eigenstates behave as statistical ensembles and eigenstate expectation values {\em already} approximate the thermal equilibrium mean. More precisely, the expectation value of any local observable $\hat{O}$, evaluated for {\em any} eigenstate $\left|\alpha\right\rangle $ of a chaotic system, is
approximately equal to its microcanonical mean over the pertinent energy shell,
\begin{equation}
\left\langle \alpha\left|\hat{O}\right|\alpha\right\rangle \approx\overline{\left\langle \alpha\left|\hat{O}\right|\alpha\right\rangle }\equiv\frac{\Delta_{\alpha}}{\Delta_{\mathrm{MC}}}\sum_{\alpha'\in\mathrm{MC}(E_{\alpha})}\left\langle \alpha'\left|\hat{O}\right|\alpha'\right\rangle ,\label{eq:ETH}
\end{equation}
where $E_{\alpha}$ is the eigenstate energy, $\alpha\in\mathrm{MC}(E)$
means that $|E_{\alpha}-E|<\Delta_{MC}/2$, $\Delta_{\mathrm{MC}}$
is the microcanonical shell width, and $\Delta_{\alpha}$ is the average
distance between the adjacent $E_{\alpha'}$ in the vicinity of $E_{\alpha}$.

Equation \eqref{eq:ETH} provides an equilibrium state that
is independent of the initial state details,
 but does not provide the equilibrium state properties.
For a low-density gas
of interacting particles in a flat potential the equilibrium state
agrees with the microcanonical ensemble for an ideal gas, as proven
in \cite{srednicki1994} on the basis of
the Berry conjecture \cite{berry1977} that each eigenfunction appears to be a superposition of plane waves with random phases and Gaussian random amplitudes, but fixed wavelength. In the thermodynamic limit, where the number
of particles and the system's volume are increased while keeping a fixed particle density,
the microcanonical ensemble provides
the Fermi-Dirac (FD) or Bose-Einstein (BE) momentum distributions  for the respective permutation symmetry, with the standard
relation between the temperature and the total gas energy, which is
equal to the eigenstate energy. Such distributions were also obtained
for interacting Fermi \cite{horoi1995,flambaum1997a} and Bose \cite{borgonovi2017}
systems close to quantum degeneracy, by appropriately shifting the microcanonical shell energies of the non-interacting system, effectively changing its temperature.

The eigenstates of the non-integrable system are superpositions of the integrable-system eigenstates. The number of
principal components (NPC) $\mathcal{N}_{\mathrm{PC}}$ estimates (for each exact eigenstate) the number of contributing integrable-system eigenstates (see App. \ref{sec:ChaoticProp}).
ETH means that the eigenstate to eigenstate fluctuations of expectation
values within any chaotic microcanonical shell are suppressed. In certain situations \cite{neuenhahn2012,yurovsky2023},
the fluctuation variance is inversely-proportional to  $\mathcal{N}_{\mathrm{PC}}$.
Thus, ETH typically implies large NPC, but can be practically attained when NPC is substantially smaller than the
dimension  $\mathcal{N}_{\mathrm{HS}}$ of the Hilbert space (which can be also constrained due to possible conservation laws).  NPC approaches a large fraction  (limited by $1/3$ for time-reversible systems \cite{kota}  or $1/2$ for time-irreversible ones \cite{truong2016}) of  $\mathcal{N}_{\mathrm{HS}}$ only in the regime of quantum ergodicity \cite{altshuler1997}.

In this work we demonstrate substantial qualitative deviations from the FD and BE distributions
in certain strongly-interacting, quantum-ergodic systems. These deviations go beyond any mapping between the microcanonical shells of the interacting- and non-interacting system (see e.g. Ref. \cite{borgonovi2017}), as the distributions become non-monotonic.

Known deviations from the FD and BE distributions can be attributed to a lack of ergodicity.
For example, quasi-integrable systems remember their initial state, as observed in experiments
with quantum Newton cradles \cite{kinoshita2006,tang2018} and cold-atom
breathers \cite{dicarli2019,luo2020}. The relaxation outcomes for integrable systems are captured by generalized Gibbs ensembles \cite{rigol2007} that account for additional integrals of motion. Incompletely chaotic systems \cite{yurovsky2011,olshanii2012} with a small number of degrees of freedom keep certain memory of their initial states. In many-body systems, eigenstate thermalization can
also be prevented by many-body localization (MBL) \cite{altshuler1997,abanin2019},
vanishing in the thermodynamic limit (see also, e.g., \cite{kiefer2020,kiefer2021,sierant2021,sierant2022}).
Even if ergodicity exists and eigenstate thermalization does take place, the distributions can
deviate from the FD and BE ones due to moderate numbers of degrees of freedom in mesoscopic systems \cite{vardi2024}. This effect, however, vanishes in large systems. Unlike all the above mechanisms, our results here are obtained in the quantum ergodic regime and survive in large systems.

\section{The lattice models} \label{sec:models}
We find  eigenstates of two lattice models by direct numerical diagonalization, allowed up to the Hilbert space dimension $\mathcal{N}_{\mathrm{HS}}\lesssim 2\times 10^4$.  Throughout this manuscript, all  energies are measured in units of the lattice
hopping energy. In the first, two-dimensional (2D)
Fermi-Hubbard (FH) model, $N$ spin-polarized fermions have nearest-neighbor
interactions with the strengths $V$ (see App. \ref{sec:2Dlatt}). This model
includes hoppings with simultaneous change of $l_{x}$
and $l_{y}$, which label sites of the $L_{x}\times L_{y}$
lattice. This, together with twisted-periodic boundary conditions,
allow us to remove degeneracies of the many-body non-interacting particle
eigenstates. The total number of one-body (1B) states in this model
is $L=L_{x}L_{y}$. Due to the spatial homogeneity of this model, we consider separately each sector with the given total momentum which contains
$\mathcal{N}_{\mathrm{HS}}\approx(L-1)!/(N!(L-N)!)$ eigenstates. The results below are obtained for $N=6$ particles in the $6\times 5$ lattice ($L=30$) and the total momentum $x$ and $y$ components $3$ and $2$, respectively. In this case,
$\mathcal{N}_{\mathrm{HS}}=19811$.

The second model  is a one-dimensional (1D) Bose-Hubbard (BH) chain
with $N$ spinless bosons in $L$ sites, with on-site interactions of strength
$V$  and hard wall boundaries  (see App. \ref{sec:1DBH}). Parity symmetry is broken
by adding a random disorder/bias potential of order $\le 0.05$ (see also \cite{chakrabarti2024}).  The resulting Hilbert space
dimension for the bosonic system is
$\mathcal{N}_{\mathrm{HS}}=(N+L-1)!/(N!(L-1)!)$. The system with $N=10$ particles in $L=8$ sites, considered here, has
$\mathcal{N}_{\mathrm{HS}}=19448$.

Analyzing the chaotic system properties, we have to
compare them to ones of the closest integrable system. For this purpose,
we use corresponding systems of non-interacting particles. Their symmetric
or anti-symmetric many-body eigenfunctions --- the orbital Fock states $\left|n\right\rangle =\left|n_{1}\ldots n_{L}\right\rangle $
--- have the eigenenergies $E_{n}=\sum_{k}n_{k}\varepsilon_{k}$.
Here $n_{k}$ are occupations of the 1B orbitals, labeled in increasing
order of their eigenenergies $\varepsilon_{k}$. Subtractions the average expectation values of interactions  from the interacting particle Hamiltonians  (see Apps. \ref{sec:2Dlatt} and \ref{sec:1DBH})  leads  to a substantial overlap of the non-interacting and interacting  spectra $\{E_{n}\}$  and  $\{E_{\alpha}\}$ .
\begin{figure}
\includegraphics[clip,width=3.5in]{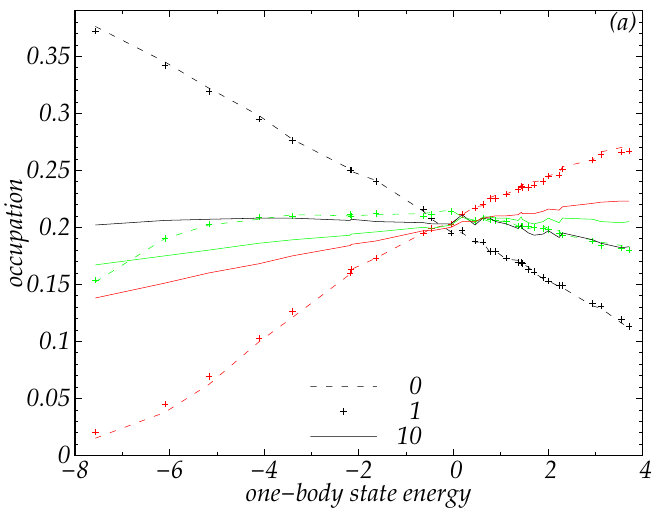}

\includegraphics[width=3.5in]{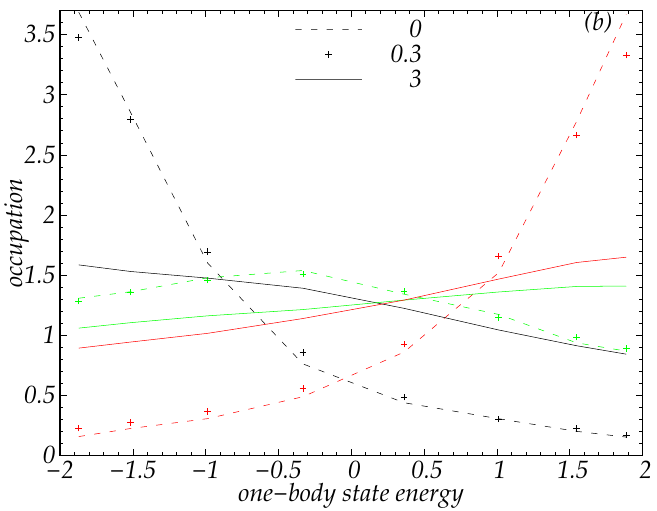}

\caption{1B orbital occupations for (a) the FH model with the interaction strengths $V=0$ (dashed lines),
$V=1$ (pluses), and $V=10$ (solid lines), averaged over microcanonical
shells with the mean energies $-6$ (black), $0$ (green), and $6$ (red) and
(b) the 1D BH model with the interaction
strengths $V=0$ (dashed lines), $V=0.3$(pluses), and $V=3$ (solid
lines), averaged over microcanonical shells with the mean energies
$-1.2$ (black), $-0.17$ (green), and $1.2$ (red). \label{fig:One-body-orbital-occupations}}
\end{figure}

\section{Exact diagonalization results}
For sufficiently strong interaction, both models become chaotic. For the FH model at
$V=1$ the ratio of two consecutive level spacings \cite{oganesyan2007},
averaged over the energy spectrum  (see App. \ref{sec:ChaoticProp}) increases
to $\left\langle r\right\rangle \approx0.525$ (cf. $\left\langle r\right\rangle \approx0.536$ \cite{atas2013}
for the Wigner-Dyson ensemble of Gaussian orthogonal matrices, describing
completely-chaotic systems). The chaotic behavior is confirmed also
by suppression of eigenstate-to-eigenstate fluctuations of
the observable expectation values (see App. \ref{sec:ChaoticProp}). Their variances are
reduced by two orders of magnitude. Another criterion of chaoticity,
NPC, increased to $5\times10^{2}$ (see App. \ref{sec:ChaoticProp}). The chaoticity of the BH
model is determined by the value of $VN$. For
the BH model at $VN=3.0$ ($V=0.3$) we have $\left\langle r\right\rangle \approx0.529$,
fluctuation variances are reduced by two orders of magnitude, and
NPC is increased to $8\times10^{2}$. For these interaction strengths,
the microcanonical distributions of the 1B orbital occupations  \footnote{The orbital occupations  can be directly observed in experiments. In addition, all single-particle parameters can be expressed in terms of them.} for interacting
and non-interacting particles are very close (see Fig. \ref{fig:One-body-orbital-occupations}). This is in line with \cite{borgonovi2017}, as subtracting the average expectation values of interactions  from the interacting particle Hamiltonians is equivalent to the energy shift used there. We notice, that both interacting
and non-interacting distributions are different from the FD and BE distribution due to the small system size \cite{vardi2024}. Due to macroscopic self-trapping, the BH model
becomes integrable again at large $V$ where the site populations become effective
integrals of motion. Thus, the chaoticity parameter reduces to  $\left\langle r\right\rangle \approx0.41$
at $V=10$ (cf $\left\langle r\right\rangle \approx0.386$ for
integrable systems).

By contrast, when the interaction is increased (but remains in the chaos
region for the BH model) the microcanonical distributions for interacting
particles deviate substantially from the non-interacting ones. This is
shown in Fig. \ref{fig:One-body-orbital-occupations} for the FH model
with $V=10$, when $r\approx0.53$ and NPC increases to the value
of $6.2\times10^{3}$, about one third of $\mathcal{N}_{\mathrm{HS}}=19811$
and for the BH model with $V=3$, $r\approx0.5$, and NPC $4.5\times10^{3}$
(about one quarter of $\mathcal{N}_{\mathrm{HS}}=19448$). The selected energy shells represent the positive temperature range
$E<(N/L)\sum\varepsilon_k$ where the distribution is monotonically decreasing with single-particle energy, the negative temperature range
$E>(N/L)\sum\varepsilon_k$ where it is monotonically increasing, and the transition region between them where it has a maximum for our small systems in the case of weak interactions  \cite{vardi2024}. Unlike the weak-interaction case, the
interaction-energy shift implicit in our comparison (equivalent to Ref.\cite{borgonovi2017}) does not capture the deviation, nor does any other energy shift, as the interacting and non-interacting distributions have different curvatures and are thus {\em qualitatively} different. As the eigenstate-to-eigenstate fluctuations are suppressed, the distributions for individual eigenstates deviate too.
\section{Non-monotonic distributions}
The effect can be explained in the following way.  Consider an observable
$\hat{O}$ that commutes with the Hamiltonian of non-interacting particles,
such that $\hat{O}\left|n\right\rangle =O_{n}\left|n\right\rangle $.
The microcanonical mean \eqref{eq:ETH} of its expectation value evaluated for
eigenstates of interacting particles can be expressed as:
\begin{equation}
\overline{\left\langle \alpha\left|\hat{O}\right|\alpha\right\rangle }=\sum_{n}\Delta_{\alpha}W(E,E_{n})O_{n}\label{eq:meanInt}
\end{equation}
in terms of the local density of states (LDOS), or strength function
\cite{kota}
\begin{equation}
W(E,E_{n})=\frac{1}{\Delta_{\mathrm{MC}}}\sum_{\alpha\in\mathrm{MC}(E)}\left|\left\langle \alpha|n\right\rangle \right|^{2}\label{eq:LDOS}
\end{equation}
{[}see \eqref{eq:ETH}{]}. The LDOS is generally a flat function of energies.
If its energy span $\Gamma$ substantially exceeds $\Delta_{\mathrm{MC}}$,
$O_{n}$ in \eqref{eq:meanInt} is effectively averaged and can be
approximated by its microcanonical mean
\begin{equation}
\overline{O}(E)=\frac{\Delta_{n}(E)}{\Delta_{\mathrm{MC}}}\sum_{n\in\mathrm{MC}(E)}O_{n},
\label{Obar}
\end{equation}
where $\Delta_{n}(E)$ is the average distance between neighboring $E_{n}$
in the vicinity of $E$. Further, as $\Gamma$ substantially exceeds
$\Delta_{n}$, approximating summation in \eqref{eq:meanInt} by integration,
we get
\begin{equation}
\overline{\left\langle \alpha\left|\hat{O}\right|\alpha\right\rangle }\approx\int_{E_{\mathrm{min}}}^{E_{\mathrm{max}}}\frac{dE'}{\Delta_{n}(E')}\Delta_{\alpha}(E)W(E,E')\overline{O}(E'),\label{eq:meanIntmean}
\end{equation}
where $E_{\mathrm{min}}$ and $E_{\mathrm{max}}$  define the support of the non-interacting
system's spectrum $E_{\{n\}}$.

Consider a particular case of the 1B orbital occupation operator $\hat{N}_{k}\left|n\right\rangle =n_{k}\left|n\right\rangle $, where $k$  labels the 1B orbitals  in increasing order of their eigenenergies $\varepsilon_{k}$  (see Sec. \ref{sec:models} and Apps. \ref{sec:2Dlatt} and \ref{sec:1DBH}).
The microcanonical distribution of the orbital occupations for non-interacting particles
$\overline{N_{k}}(E)$ is given by Eq. \eqref{Obar}. The  shape of $\overline{N_{k}}(E)$  depends on the mean shell energy $E$ (see Fig.
\ref{fig:One-body-orbital-occupations}). If $W(E,E')$ vanishes when
$|E-E'|>\Gamma$ and $\Gamma$ is small with respect to the energy scale on which the microcanonical distribution $\overline{N_{k}}(E)$ varies, we can approximate the microcanonical distribution for interacting particles $\overline{N_{k}^{int}}(E)\equiv\overline{\left\langle \alpha\left|\hat{N}_{k}\right|\alpha\right\rangle }$ [see Eq. \eqref{eq:meanIntmean})] as  $\overline{N_{k}^{int}}(E)\approx\overline{N_{k}}(E)$, thus justifying the equivalence between the occupation statistics of the interacting and non-interacting systems.
However, if $\Gamma$ exceeds this scale, the interacting-system's occupation distribution $\overline{N_{k}^{int}}(E)$ can mix non-interacting distributions $\overline{N_{k}}(E)$ of different shape and be different
from any individual non-interacting microcanonical distribution $\overline{N_{k}}(E)$. In Fig. \ref{fig:One-body-orbital-occupations}, mixing of increasing and decreasing $\overline{N_{k}}(E)$ leads to near uniform   $\overline{N_{k}^{int}}(E)$ .

The exact diagonalization method is applicable only to small numbers
of particles and lattice sites when the microcanonical distribution
of the orbital occupations $\overline{N_{k}}(E)$ is different from
the FD and BE distributions \cite{vardi2024}. However, for large
numbers of non-interacting particles the microcanonical and canonical thermodynamic means become equivalent
(although the fluctuations in different ensembles can be nonequivalent even in
the thermodynamic limit \cite{christensen2021,crisanti2024}).
Then, the microcanonical  occupations of the orbitals
$\overline{N_{k}}(E)$ are precisely given by the FD or BE distributions
\begin{equation}
\overline{N_{k}}(E)=\left(e^{(\varepsilon_{k}-\mu)/T}\pm1\right)^{-1},\label{eq:Nkbar}
\end{equation}
where the chemical potential $\mu$ and temperature $T$ are solutions
to the system of equations $\sum_{k}\overline{N_{k}}(E)=N$ and $\sum_{k}\varepsilon_{k}\overline{N_{k}}(E)=E$.
If $\varepsilon_{k}$ is restricted both from below and above, $T$
can be either positive or negative, corresponding to occupation distributions
which decrease or increase, respectively, with the orbital energy.
The summation over $k$ in this system can be replaced by integration
over the orbital energy. Then $\mu$ and $T$ will depend on the particle
density $\tilde{N}=N/L$ and energy density $\tilde{E}=E/L$.

While finding the exact LDOS by direct diagonalization is not possible for large systems, in the case of strong interactions, it can be approximated
by the Gaussian shape (see \cite{kota})
\begin{equation}
W(E,E_{n})\approx C(E)\frac{\Delta_{n}(E_n)}{\Delta_{\alpha}(E)}\exp(-(E-E_{n})^{2}/\Gamma^{2}).\label{eq:WGauss}
\end{equation}
where $\Delta_{n}(E_n)$ is taken in the vicinity of $E_{n}$ and the normalization
factor $C(E)$ is determined by
\begin{equation}
1/C(E)=\int_{E_{\mathrm{min}}}^{E_{\mathrm{max}}}\exp(-(E-E')^{2}/\Gamma^{2})dE'.\label{eq:normW}
\end{equation}
\begin{figure}
\includegraphics[width=3.5in]{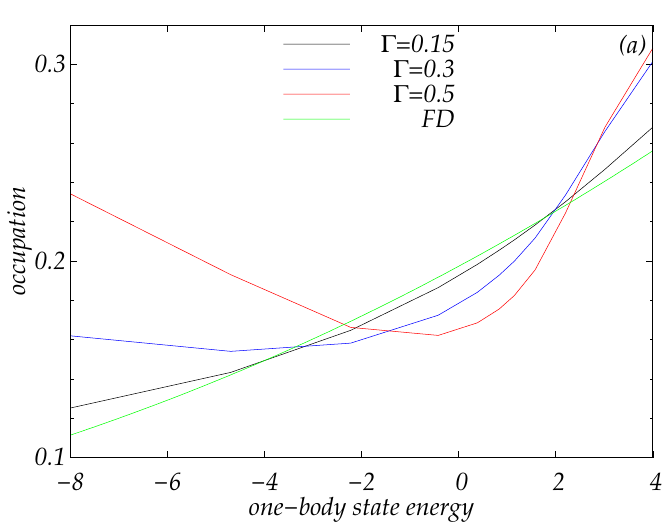}
\includegraphics[width=3.5in]{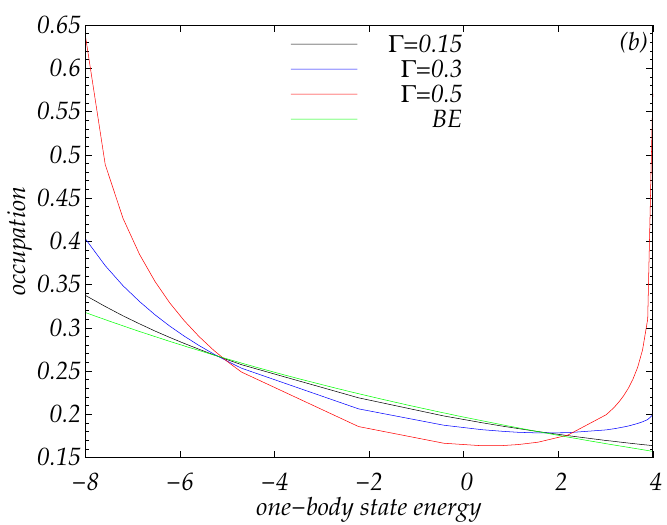}
\includegraphics[width=3.5in]{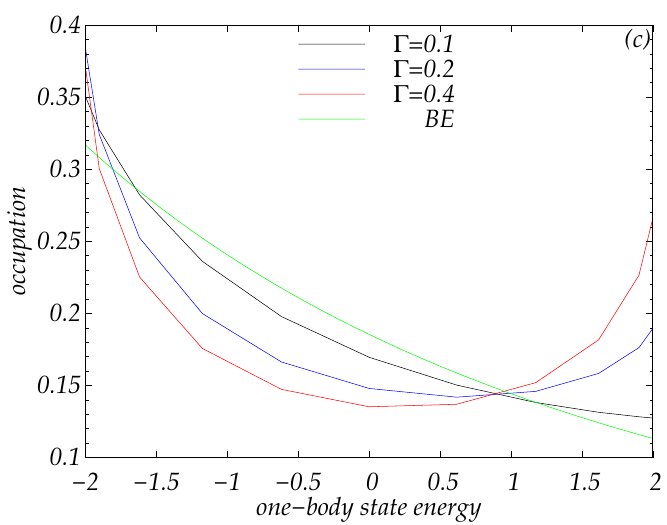}
\caption{1B orbital occupations for different $\tilde{\Gamma}$. (a) The FH
model with $\tilde{N}=0.2$ and $\tilde{E}=0.1$. (b) The 2D BH model
with $\tilde{N}=0.2$ and $\tilde{E}=-0.1$. (c) The 1D BH model with
$\tilde{N}=0.2$ and $\tilde{E}=-0.1$. The green lines show the FD
or BE distributions corresponding to $\tilde{E}$. \label{fig:largesys}}
\end{figure}
The Gaussian shape \eqref{eq:WGauss} approximates the LDOS  with good accuracy even for
systems of small size (see App. \ref{sec:LDOS}). It should be stressed,
that the agreement can be provided by the factor $\Delta_{n}(E_n)$ even if the spectrum is substantially inhomogeneous. The resulting
distributions, calculated with Eqs. \eqref{eq:meanIntmean}, \eqref{eq:Nkbar},
\eqref{eq:WGauss}, and \eqref{eq:normW} depend on scaled widths
$\tilde{\Gamma}=\Gamma/L$. In addition to the 2D FH and 1D BH model,
treated above using exact diagonalization, we consider also the 2D
BH model with the same 1B Hamiltonian as the 2D FH one  (see App. \ref{sec:2Dlatt}).
Figure \ref{fig:largesys} shows the obtained distributions for the Gaussian width
$\Gamma$ (increasing with the interaction strength $V$)  covering both eigenstates corresponding to positive and
negative temperature, or, respectively, to the decreasing and increasing
FD or BE distributions. The resulting distributions for the interacting system have clearly pronounced minima. This is provided by the positive second derivative $d^2\overline{N_{k}}(E)/d\varepsilon^2_{k}>0$ for the FD and BE distributions \eqref{eq:Nkbar} when $(\varepsilon_{k}-\mu)/T>0$. In contrast, for the few-mode systems of Fig. \ref{fig:One-body-orbital-occupations}, the minimum does not appear due to a maximum of  $\overline{N_{k}}(E)$ ($d^2\overline{N_{k}}(E)/d\varepsilon^2_{k}<0$) with $E\approx0$, and the maximum vanishes in $\overline{N_{k}^{int}(}E)$ due to strong interactions.

Since $\mathcal{N}_{\mathrm{PC}}\sim\Gamma/\Delta_{n}$
and $\mathcal{N}_{\mathrm{HS}}\sim(E_{\mathrm{max}}-E_{\mathrm{min}})/\Delta_{n},$ the
ratio of NPC to the Hilbert space dimension can be estimated as $\mathcal{N}_{\mathrm{PC}}/\mathcal{N}_{\mathrm{HS}}\sim\Gamma/(E_{\mathrm{max}}-E_{\mathrm{min}})\approx\tilde{\Gamma}/(\tilde{N}\Delta\varepsilon)$,
where the range of $\varepsilon_{k}$ variation $\Delta\varepsilon$
is 12 for the 2D and 4 for the 1D models. Then, in Fig. \ref{fig:largesys},
$\tilde{\Gamma}=0.5$ corresponds to $\mathcal{N}_{\mathrm{PC}}/\mathcal{N}_{\mathrm{HS}}\sim0.2$
for the 2D models and $\tilde{\Gamma}=0.4$ corresponds to $\mathcal{N}_{\mathrm{PC}}/\mathcal{N}_{\mathrm{HS}}\sim0.5$
for the 1D BH. These high participation fractions indicate quantum ergodicity \cite{altshuler1997}.

\section{Discussion}
It should be realized that Fig. \ref{fig:One-body-orbital-occupations} and Fig. \ref{fig:largesys} show the same effect, namely the mixing of occupation distributions obtained from different microcanonical shells of the non-interacting system, due to a broadened LDOS. However, whereas in Fig. \ref{fig:largesys}  the mean natural-mode occupations within microcanonical shells of the non-interacting system take the ubiquitous FD or BE forms  \eqref{eq:Nkbar}, the corresponding distributions for few-mode mesoscopic system that can be numerically diagonalized as in Fig. \ref{fig:One-body-orbital-occupations}, have non-universal and sometimes non-monotonic structure in themselves \cite{vardi2024}. The resulting  strong-interaction occupation distribution is hence qualitatively different in the two cases. While numerical complexity precludes direct diagonalization for large systems, the non-monotonic distributions in Fig. \ref{fig:largesys}  are the inevitable result  of the well-known FD or BE noninteracting shell distributions and the Gaussian lineshape of the LDOS \eqref{eq:WGauss}  that is confirmed in numerical calculations \cite{kota} and extends also to smaller systems (see App.\ref{sec:LDOS}). Moreover, even this particular lineshape is not necessary, as non-monotonic distributions eventually appear for LDOS of any shape as long as it is sufficiently broadened by strong interactions.

In the MBL literature, the proportionality of NPC
to Hilbert space dimension is used as an attribute of delocalization, distinguishing
extended eigenstates from localized eigenstates. This property was reported for eigenstates
in Heisenberg \cite{luitz2015} and XXZ \cite{luitz2020} spin chains,  the Bose-Hubbard
model \cite{pausch2021}, and
the Jaynes-Cummings-Hubbard system \cite{ma2022}. Then, the ratio
$\tilde{\Gamma}/\tilde{E}\sim\tilde{N}\Delta\varepsilon\mathcal{N}_{\mathrm{PC}}/(\mathcal{N}_{\mathrm{HS}}\tilde{E})$
should remain unchanged for extended eigenstates in the thermodynamic
limit $N\rightarrow\infty$, while $\tilde{N}=\mathrm{const}$ and
$\tilde{E}=\mathrm{const}$. As a result, the distribution deviations
from the FD and BE ones may survive in the thermodynamic limit.

For bosonic systems, there is a clear classical mean-field limit wherein the field operators are replaced by $c$-numbers and their amplitudes and phases serve as conjugate action-angle canonical variables. The observed broadening of the LDOS may then be viewed as resulting from the interaction-induced deformation of the energy shells within the classical phasespace. For the boson models discussed here, there is good quantum-classical correspondence in the sense that mean occupations agree well with semiclassical averages over the pertinent shells (see \cite{vardi2024,khripkov2018}) and the mean LDOS corresponds to the overlap of the classical shell of the non-interacting system with each of the interacting system's energy shells. While weak interactions only slightly shift the non-interacting shells, strong interactions deform them substantially: The non-interacting shell overlaps with many interacting shells,  resulting in the broadening of the LDOS.

We reemphasize that the above deviations from the FD or BE distributions are quite different from those
observed in \cite{borgonovi2017}. The distributions  presented in Fig. 2 therein are monotonic and agree perfectly with the BE distributions for the temperatures corresponding to the dressed energies. The dressed energy, given by  Eq. (12) in \cite{borgonovi2017}, is shifted by the average expectation value of the interactions.  As noted earlier, this prescription is implicit in all our calculations, as the average
expectation values of the interactions are subtracted from the interacting-system
eigenenergies (see Apps. \ref{sec:2Dlatt} and  \ref{sec:1DBH}). The deviations we observe are more
profound than this simple energy shift. Our main point is that the interactions
mix different microcanonical shells of the non-interacting system, so that the
microcanonical occupation means over the interacting system's energy shell do
not match \textit{any} of the corresponding microcanonical means over non-interacting
shells. This voids all energy shift prescriptions, including that of \cite{borgonovi2017}. The non-monotonic distributions obtained here are a result of this mixing of shells with positive- and negative temperature
and can not be reduced to a change of temperature.
Mixing of different shells was already considered in classical superstatistics \cite{beck2003} and thermodynamics of small systems \cite{davis2022}, but with no relation to quantum eigenstates and non-monotonic distributions.

\section{Conclusion}
Orbital population distributions in eigenstates of strongly-interacting
many-body systems can be non-monotonic and hence qualitatively deviate from the FD and BE distributions
while the eigenstates are chaotic  and thermalize. Unlike previously observed non-monotonic occupation distributions in weakly-interacting mesoscopic systems \cite{vardi2024}, this strong-interaction effect appears due to the mixing of microcanonical shells with temperatures of opposite sign and survives in large systems. The distribution deviations may be observed experimentally with cold atoms in optical lattices.
\section*{Acknowledgement}
VY and AV acknowledge support from the NSF through a grant for ITAMP at Harvard University.


\begin{appendix}
\numberwithin{equation}{section}

\section{Two-dimensional lattice models}
\label{sec:2Dlatt}

The Fermi-Hubbard (FH) model on a two-dimensional (2D) lattice has
the Hamiltonian
\begin{align}
\hat{H}_{F}  =&-\sum_{l_{x}=1}^{L_{x}}\sum_{l_{y}=1}^{L_{y}}\sum_{\delta_{x}=-1}^{1}\sum_{\delta_{y}=-1}^{1}\left(1-\delta_{\delta_{x}0}\delta_{\delta_{y}0}\right)\hat{a}_{l_{x}l_{y}}^{\dagger}\hat{a}_{l_{x}+\delta_{x}l_{y}+\delta_{y}}
\nonumber
\\
&+\hat{V}_{F}-\bar{V},\label{eq:FH}\\
\hat{V}_{F}  =&V\sum_{l_{x}=1}^{L_{x}}\sum_{l_{y}=1}^{L_{y}}\Bigl(\sum_{\delta_{x}=\pm1}\hat{a}_{l_{x}l_{y}}^{\dagger}\hat{a}_{l_{x}+\delta_{x}l_{y}}^{\dagger}\hat{a}_{l_{x}l_{y}}\hat{a}_{l_{x}+\delta_{x}l_{y}}
\nonumber
\\
&+\sum_{\delta_{y}=\pm1}\hat{a}_{l_{x}l_{y}}^{\dagger}\hat{a}_{l_{x}l_{y}+\delta_{y}}^{\dagger}\hat{a}_{l_{x}l_{y}}\hat{a}_{l_{x}l_{y}+\delta_{y}}\Bigr),
\label{eq:VF}
\end{align}
where $V$ is the nearest-neighbor interaction strength, $\bar{V}$ is  the average expectation values of the interactions $\hat{V}_{F}$ (see below), the hopping
energy is used as the energy unit, $\hat{a}_{l_{x}l_{y}}$are annihilation
operators of spin-polarized fermions, and $l_{x}$, $l_{y}$ specify
location on the the $L_{x}\times L_{y}$ lattice. Outside the square
$1\leq l_{x}\leq L_{x}$, $1\leq l_{y}\leq L_{y}$, the field operators
are defined by the twisted periodic boundary conditions $\hat{a}_{l_{x}+L_{x}l_{y}}=e^{i\chi_{x}}\hat{a}_{l_{x}l_{y}}$,
$\hat{a}_{l_{x}l_{y}+L_{y}}=e^{i\chi_{y}}\hat{a}_{l_{x}l_{y}}$. The
phase changes $\chi_{x}=(1+\sqrt{5})/2$ (the golden ratio) and $\chi_{y}=e/2$
are used in the present calculations. The 1B orbitals are plane
waves with the momentum components
\begin{align}
p_{x}=\frac{2\pi m_{x}+\chi_{x}}{L_{x}}\quad(1\leq m_{x}\leq L_{x})
\nonumber
\\
\label{eq:pxy}
\\
p_{y}=\frac{2\pi m_{y}+\chi_{y}}{L_{y}}\quad(1\leq m_{y}\leq L_{y}),\nonumber
\end{align}
where $m_{x}$ and $m_{y}$ are integers. The orbital energies are
expressed as
\begin{equation}
\varepsilon_{2D}(p_{x},p_{y})=-2\cos p_{x}-2\cos p_{y}-4\cos p_{x}\cos p_{y}.
\end{equation}
The $k$th orbital momentum components $p_{x}(k)$ and $p_{y}(k)$ are chosen such that the orbitals are labeled in increasing order of their eigenenergies
$\varepsilon_{k}=\varepsilon_{2D}(p_{x}(k),p_{y}(k))$.
In the limit of the large $L_{x,y}$ the number of the orbitals with
energies below $\varepsilon$, $k(\varepsilon)$, can be approximated by
\begin{equation}
\frac{k(\varepsilon)}{L}\approx\frac{1}{(2\pi)^{2}}\int_{0}^{2\pi}dp_{x}\int_{0}^{2\pi}dp_{y}\vartheta(\varepsilon-\varepsilon_{2D}(p_{x},p_{y})),
\end{equation}
where summation over $m_{x,y}$ is approximated by integration over
$p_{x,y}$, $L=L_{x}L_{y}$ is the total number of orbitals, and $\vartheta$
is the Heaviside step function. Inversion of $k(\varepsilon)$ allows
us to express $\varepsilon_{k}=\varepsilon(k/L)$ in terms of lattice-size
independent function $\varepsilon(\tilde{k)}$ which increases with
$\tilde{k}$ from $\varepsilon(0)=-8$ to $\varepsilon(1)=4$ (see
Fig. \ref{fig:en1b2d}).
\begin{figure}
\includegraphics[width=3.25in]{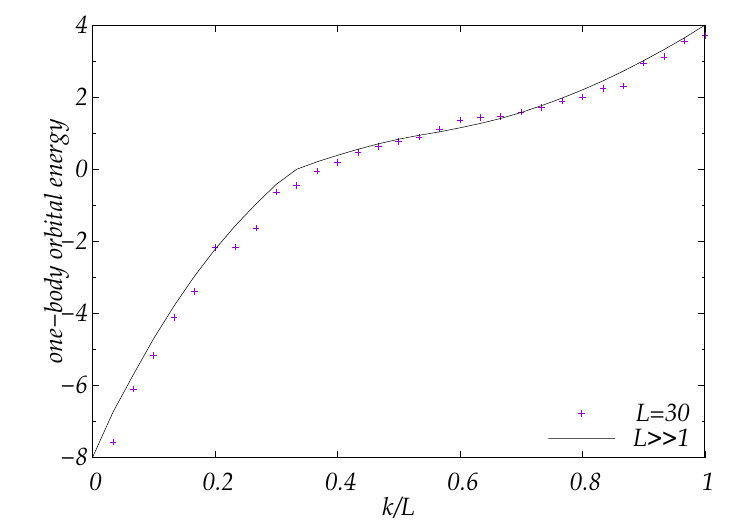}
\caption{One-body orbital energy as a function of the orbital label $k$ for
small and large FH models.\label{fig:en1b2d}}
\end{figure}

The eigenstates of the Hamiltonian \eqref{eq:FH}  with $V=0$ are thus the orbital Fock states $|n\rangle=|n_1,...n_L\rangle$ where $0\leq n_k\leq 1$ is the integer occupation of the $k$-th orbital, and $\sum_{k=1}^L n_k=N$. Due to spatial homogeneity of the Hamiltonian \eqref{eq:FH}, we consider separately each segment with given total momentum components $P_x$ and $P_y$, such that  $\sum_{k=1}^L n_k p_{x,y}(k)=P_{x,y}$.  The orbital Fock states for each segment constitute a $\mathcal{N}_{\mathrm{HS}}\approx(L-1)!/(N!(L-N)!)$ dimensional complete basis for the many-body Hilbert space with given   $P_x$ and $P_y$. Representing the full Hamiltonian in this basis and diagonalizing, we obtain the exact many-fermion eigenstates $|\alpha\rangle$.

The average expectation values of interactions
$\bar{V}=\sum_{\alpha}\left\langle \alpha\left|\hat{V}_F\right|\alpha\right\rangle /\mathcal{N}_{\mathrm{HS}}$ is subtracted in the Hamiltonian  \eqref{eq:FH} in order to provide a substantial overlap between the non-interacting and interacting  spectra $\{E_{n}\}$  and  $\{E_{\alpha}\}$. Due to completeness of the set $\left|\alpha\right\rangle $, we have
\begin{equation}
 \bar{V}=\frac{1}{\mathcal{N}_{\mathrm{HS}}}\sum_{n}\left\langle n\left|\hat{V}_F\right|n\right\rangle ,
 \label{eq:barVn}
\end{equation}
where diagonal matrix elements of the interaction  \eqref{eq:VF} can be expressed as
\begin{align}
 \left\langle n\left|\hat{V}_F\right|n\right\rangle=\frac{4V}{L}
 \sum_{k<k'} & n_k n_k'\Bigl( \sin^2\frac{p_x(k)-p_x(k')}{2}
 \nonumber
 \\
 &+\sin^2\frac{p_y(k)-p_y(k')}{2}\Bigr).
 \label{eq:Vnn}
\end{align}
As all orbitals are presented unbiasedly in the set $\{|n\rangle\}$, we can approximate the average over the Hilbert space in \eqref{eq:barVn} by the average over $p_{x}(k)$ and $p_{y}(k)$, i.e, replace squared sines in \eqref{eq:Vnn} by $1/2$. As a result, we get
\begin{equation}
\bar{V}\approx2N(N-1)\frac{V}{L}.
\end{equation}
This approximate value will be valid as well for the average over each microcanonical interval, where the orbitals are presented unbiasedly. The stretching of the  interacting  spectrum   $\{E_{\alpha}\}$ in comparison with the non-interacting one $\{E_{n}\}$ is related to the level repulsion, which is beyond the first order effect in $\hat{V}_F$.

The expectation values of the orbital occupations $\langle \alpha|\hat N_k|\alpha\rangle$ for each of the $\mathcal{N}_{\mathrm{HS}}$ eigenvalues are calculated with   $\sum_{k=1}^L n_k m_{x}(k)=3$, $\sum_{k=1}^L n_k m_{y}(k)=2$, and $\mathcal{N}_{\mathrm{HS}}=19811$ for $N=6$ particles in $L=30$ sites of the $6\times 5$ lattice.

We also consider a 2D Bose-Hubbard (BH) model of the large system
size. It has the Hamiltonian \eqref{eq:FH} where $\hat{a}_{l_{x}l_{y}}$ are
annihilation operators of spinless bosons and $\hat{V}_{F}$
is replaced by local interactions. The 1B Hamiltonian, orbitals, and
$\varepsilon_{k}$ for this model are the same as for the 2D FH one.

\section{One-dimensional Bose-Hubbard model}
\label{sec:1DBH}
The tight binding bosonic Hamiltonian on a one-dimensional (1D) lattice (in units of the hopping rate) reads,
\begin{equation}
{\hat H}_{B}=-\sum_{l,m=1}^L {\hat b}_l^\dag J_{lm}{\hat b}_m+\frac{1}{2} V \sum_{l=1}^L \hat n_l(\hat n_l-1)-\bar{V},
\end{equation}
where $l=1,...,L$ is the site index, $J_{lm}=J_{ml}^*$ is the hopping matrix coupling sites $l$ and $m$, $V$  is the on-site interaction strength, $\hat n_l=\hat b_l^\dag\hat b_l$ is the number of bosons at site $l$, and $\hat b_l$ are bosonic particle annihilation operators.  Throughout the manuscript we have used the Bose-Hubbard (BH) configuration $J_{l\ne m}=\delta_{l,m\pm1}$ with hard wall boundaries, i.e. a linear chain of $L$ sites. For this configuration, the dynamical behavior of the system, e.g. its degree of chaoticity, is set by the dimensionless interaction parameter $u=VN$. In order to remove the remaining parity symmetry and increase chaoticity, we have introduced a weak random 'disorder' on-site potential $J_{l,l}={\rm rnd}[-0.05,0.05]$.

The 1B orbitals are found by diagonalizing the hopping matrix, thereby obtaining the eigenvectors $\{f_\alpha\}_{k=1,...L}$ and the orbital energies $\varepsilon_k$. Defining the bosonic mode annihilation operators $\hat c_k=\sum_l f_k(l) \hat b_l$ where $f_k(l)$ denotes the $l$-th component of the $k$-th eigenvector, we obtain the orbital number operators:
\begin{equation}
\hat N_k = \hat c_k^\dag\hat c_k=\sum_{l,m} f_k^*(l) f_k(m) \hat b_l^\dag\hat b_m
\end{equation}

The BH Hamiltonian then transforms in the orbital basis into,
\begin{equation}
{\hat H}_{B}=\sum_{k=1}^L \varepsilon_k{\hat c}_l^\dag {\hat c}_l
+{\hat V}_{B}-\bar{V},
\label{BHH2}
\end{equation}
where,
\begin{equation}
{\hat V}_{B}=\sum_{k,k',k'',k'''=1}^L u_{k,k',k'',k'''}\hat c_k^\dag\hat c_{k'}^\dag \hat c_{k''} \hat c_{k'''}
\end{equation}
and
\begin{equation}
u_{k,k',k'',k'''}=\frac{V}{2} \sum_{i=1}^L f_k^*(i) f_{k'}^*(i) f_{k''}(i) f_{k'''}(i)
\end{equation}
Note that in contrast to the FH model of the previous section,  the system is not translationally invariant. Hence there is no momentum conservation law that reduces the allowed four-wave-mixing transitions induced by the interactions between the orbitals.

The eigenstates of the Hamiltonian of Eq.~(\ref{BHH2})  with $V=0$ are thus the orbital Fock states $|n\rangle=|n_1,...n_L\rangle$ where $n_k$ is the integer occupation of the $k$-th orbital, and $\sum_{k=1}^L n_k=N$. The orbital Fock states constitute a $\mathcal{N}_{\mathrm{HS}}=(N+L-1)!/(N!(L-1)!)$ dimensional complete basis for the many-body Hilbert space (throughout the manuscript $\mathcal{N}_{\mathrm{HS}}=19448$ for $N=10$ particles in $L=8$ sites). Representing the full Hamiltonian in this basis and diagonalizing, we obtain the exact many-boson eigenstates $|\alpha\rangle$ and calculate the expectation values of the orbital occupations $\langle \alpha|\hat N_k|\alpha\rangle$ for each of the $\mathcal{N}_{\mathrm{HS}}$ eigenvalues. In this model, $\bar{V}=\overline{\left\langle \alpha\left|\hat{V}_B\right|\alpha\right\rangle }$
is the microcanonical mean of the interaction expectation value. For bosons, due to multiple orbital occupations, $\bar{V}$ is energy-dependent. Then, it is numerically calculated for each microcanonical shell.

\section{Chaotic properties}
\label{sec:ChaoticProp}

The degree of chaoticity  of a quantum system can be deduced from its level spacing statistics. One measure of the transition from the Poissonian statistics of integrable systems to the Wigner-Dyson statistics of completely chaotic systems is the ratio of consecutive level spacings
\begin{equation}
r_{\alpha}=\frac{\min(E_{\alpha+1}-E_{\alpha},E_{\alpha}-E_{\alpha-1})}{\max(E_{\alpha+1}-E_{\alpha},E_{\alpha}-E_{\alpha-1})}.
\end{equation}
averaged over the entire spectrum or over a pertinent energy shell. This criterion has been introduced in \cite{oganesyan2007}, and is very widely used as a clear evidence of integrability-chaos transition.  The value $\left\langle r\right\rangle=2\ln2-1\approx0.38629$ is indicative of Poissonian statistics, whereas $\left\langle r\right\rangle =4-2\sqrt{3}\approx0.53590$ is obtained for Wigner-Dyson GOE statistics \cite{atas2013}. In Fig. \ref{fig:Level-spacing-ratio} we present this measure as a function of the interaction strength for our model systems. The 1D BH system is integrable at weak interaction due to its near-separability and at strong interaction due to macroscopic self-trapping where site occupations become integrals of motion. In contrast, the 2D FH system does not return to integrability at high interaction strength.
\begin{figure}
\includegraphics[width=3.5in]{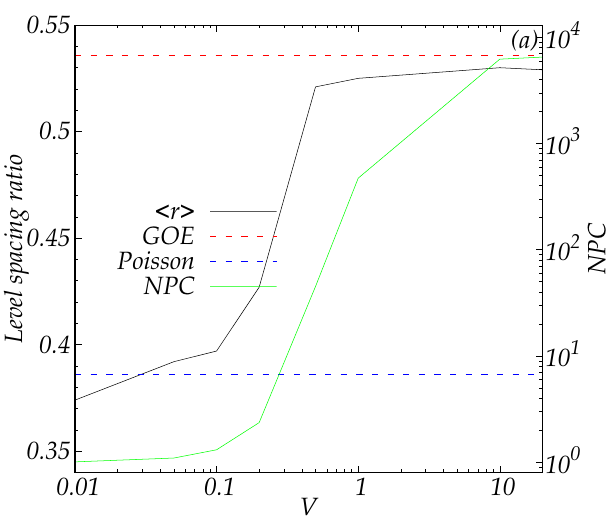}
\includegraphics[width=3.5in]{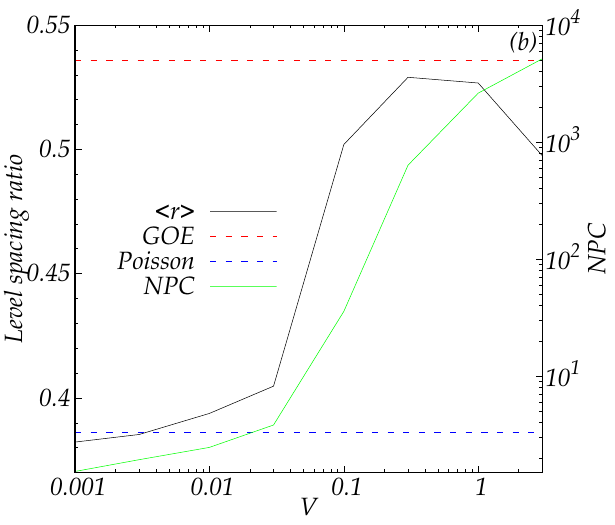}
\caption{(a) Level spacing ratio (black line) vs. interaction strength for
the FH model. Dashed lines
show $\left\langle r\right\rangle $ for the Poisson and GOE statistics.
The green line shows NPC. (b) The same for the 1D BH model. \label{fig:Level-spacing-ratio}}
\end{figure}
Figure \ref{fig:Level-spacing-ratio} presents also another characteristic of chaos --- the number of principal components (NPC), or the participation ratio,
$\mathcal{N}_{\mathrm{PC}}=\eta^{-1}$, where the inverse participation ratio is defined by
\begin{equation}
\eta=\sum_{n}\left|\left\langle n|\alpha\right\rangle \right|^{4}.
\end{equation}
In Fig. \ref{fig:Level-spacing-ratio}(a),
$\mathcal{N}_{\mathrm{PC}}\approx 6.2\times10^3$ for $V=10$ and
$\mathcal{N}_{\mathrm{PC}}\approx 6.5\times10^3$ for $V=20$, tending to the GOE limit $\mathcal{N}_{\mathrm{HS}}/3\approx 6.6\times10^3$ at
$V \rightarrow \infty$.

Note that a high NPC is a necessary but not sufficient condition for chaos, as the number of eigenstates of a non-interacting system participating in an eigenstate of the interacting system can be large even if the latter is integrable.

Chaos can also be characterized by  the eigenstate-to-eigenstate
fluctuations of the observable expectation values.
The fluctuation variances for an observable $\hat{O}$ are expressed as
\begin{equation}
 \mathrm{Var}_{\alpha}(\hat{O})= \overline{\left\langle \alpha\left|\hat{O}\right|\alpha\right\rangle ^{2}}
-  \overline{\left\langle \alpha\left|\hat{O}\right|\alpha\right\rangle }^{2}.
\end{equation}
\begin{figure}
\includegraphics[width=3.5in]{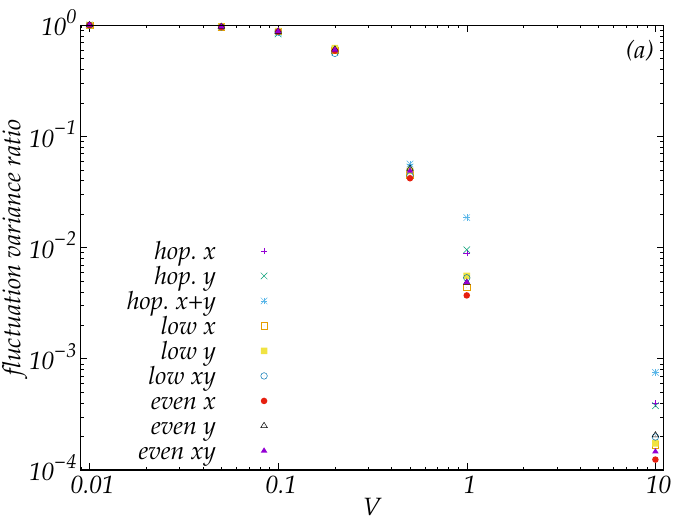}

\includegraphics[width=3.5in]{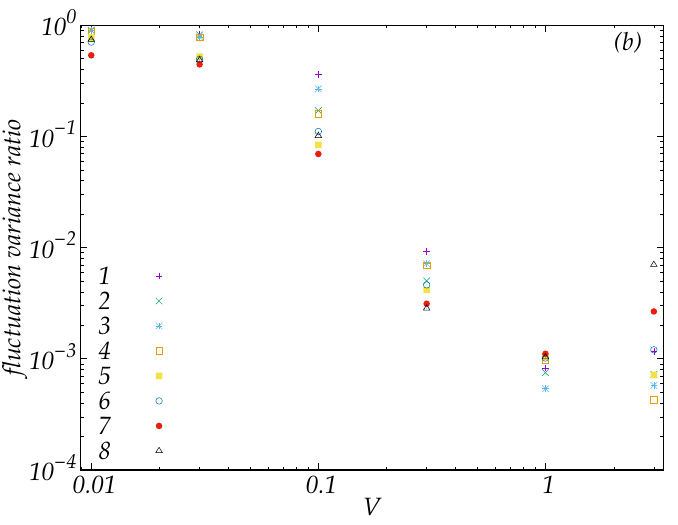}\caption{Ratio of eigenstate-to-eigenstate fluctuation variances for the non-integrable
to ones for the integrable systems eigenstates vs. interaction strength
for: (a) the FH model
averaged over the microcanonical shell with the mean energy $0$;
(b) the 1D BH model with the mean shell
energy $-5.14$. \label{fig:FlucVar}}
\end{figure}

\begin{figure}[!h]
\includegraphics[width=3.5in]{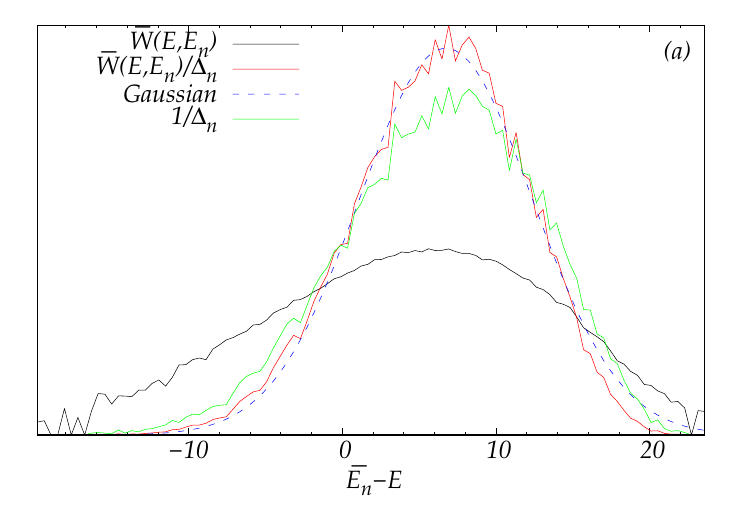}
\includegraphics[width=3.5in]{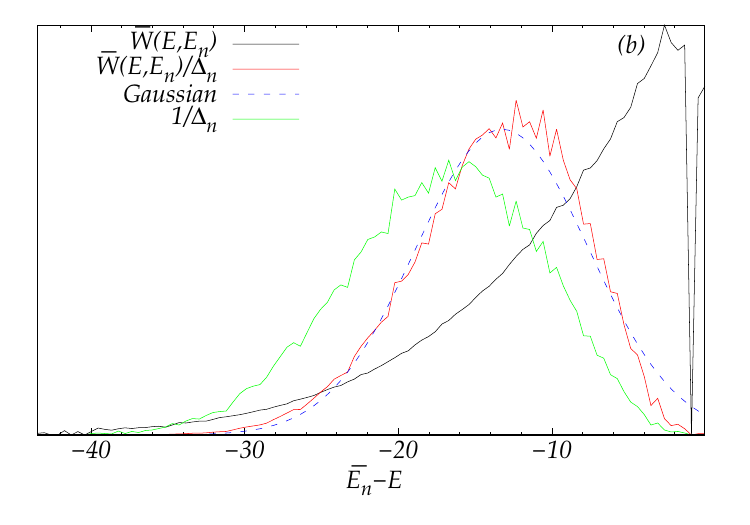}
\includegraphics[width=3.5in]{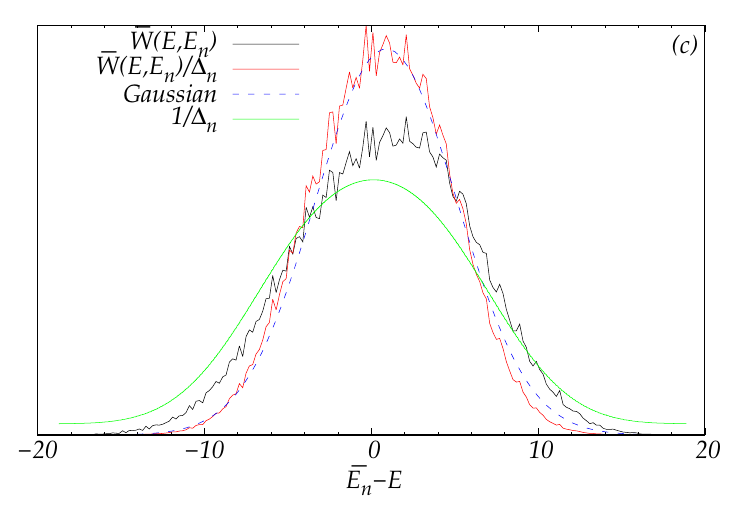}
\caption{Local density of states, averaged over the orbitals for:
(a) the FH model with $V=10$ at $E=-5.2$;
(b) the FH model with $V=10$ at $E=38.4$;
(c) the 1D BH model with $V=3$ at $E=-5.14$.
\label{fig:LDOS}}
\end{figure}
The variances are presented in Fig. \ref{fig:FlucVar}. Due to the large number of the orbitals in the FH model, we consider cumulative
observables: the total occupations of orbitals with $m_{x}<L_{x}/2$
and any $m_{y}$ in Eq. \eqref{eq:pxy} {[}low x in Fig. \ref{fig:FlucVar}(a){]},
with $m_{y}<L_{y}/2$ and any $m_{x}$ (low y), with $m_{x}<L_{x}/2$
and $m_{y}<L_{y}/2$ (low xy), with even $m_{x}$ and any $m_{y}$
(even x), with even $m_{y}$ and any $m_{x}$ (even y), and with even
$m_{x}$ and $m_{y}$ (even xy). We also consider the hopping energies
in the $x$
\[
-\sum_{l_{x}=1}^{L_{x}}\sum_{l_{y}=1}^{L_{y}}\sum_{\delta_{x}=\pm1}\hat{a}_{l_{x}l_{y}}^{\dagger}\hat{a}_{l_{x}+\delta_{x}l_{y}}
\]
and $y$
\[
-\sum_{l_{x}=1}^{L_{x}}\sum_{l_{y}=1}^{L_{y}}\sum_{\delta_{y}=\pm1}\hat{a}_{l_{x}l_{y}}^{\dagger}\hat{a}_{l_{x}l_{y}+\delta_{y}}
\]
directions, as well as the sum of these energies. For the 1D BH model,
due to the small number of orbitals, we consider the individual orbital
occupations.

\section{Local density of states}
\label{sec:LDOS}
Figure \eqref{fig:LDOS} demonstrates the local density of states
(LDOS) {[}see \eqref{eq:LDOS}{]} averaged over the orbitals
\[
\bar{W}(E,\bar{E}_{n})=\frac{\Delta_{n}(E_n)}{\Delta_{\mathrm{MC}}}\sum_{n\in\mathrm{MC}(\bar{E}_{n})}W(E,E_{n})
\]
and the averaged LDOS divided by $\Delta_{n}(E_n)$ in a comparison with
the Gaussian profiles.

In the centre of the spectrum,  both $\bar{W}(E,\bar{E}_{n})$ and $\bar{W}(E,\bar{E}_{n})/\Delta_{n}(E_n)$
can be approximated by Gaussian profiles (see Figs.  \ref{fig:LDOS}(a) and (c)). However, near the spectrum boundaries, where the spectrum is substantially inhomogeneous, only
$\bar{W}(E,\bar{E}_{n})/\Delta_{n}(E_n)$ has a Gaussian shape, but $\bar{W}(E,\bar{E}_{n})$
does not (see Fig.  \ref{fig:LDOS}(b)).
\end{appendix}

\end{document}